\documentclass[12pt]{article}
\usepackage{epsfig}
\newcommand{\mysection}{\setcounter{equation}{0}\section}

\def\beq{\begin{equation}}
\def\eeq{\end{equation}}
\def\beqa{\begin{eqnarray}}
\def\eeqa{\end{eqnarray}}

\newlength{\dinwidth} \newlength{\dinmargin}
\begin{document}
\begin {flushright}
Cavendish-HEP-04/05\\
\end {flushright} 
\vspace{2cm}
\begin{center}
{\Large \bf Next-to-next-to-leading order soft and virtual corrections for QCD,
Higgs and SUSY processes}
\end{center}
\vspace{2mm}
\begin{center}
{\large Nikolaos Kidonakis}\\
\vspace{2mm}
{\it Cavendish Laboratory\\ University of Cambridge\\
Madingley Road\\ Cambridge CB3 0HE, UK} 
\end{center}

\begin{abstract}
I review recent advances in the calculation of higher-order
soft-gluon corrections for a variety of QCD, Higgs, and SUSY processes
in hadron colliders.
A unified approach and master formulas for
next-to-next-to-leading order soft and virtual
corrections are discussed. I present some applications of the
formalism to top quark pair production at the Tevatron and the LHC, 
top production via anomalous couplings
in flavor-changing neutral-current processes, $W$ boson
hadroproduction at large transverse momentum, and charged Higgs 
production at the LHC.
\end{abstract}

\thispagestyle{empty} \newpage \setcounter{page}{2}

\mysection{Introduction}

A lot of progress in calculations of higher-order QCD corrections
has been achieved in the last few years.
Recent theoretical advances include an array of resummations,
and a few complete and several partial next-to-next-to-leading order 
(NNLO) calculations \cite{WGQCD,QCDSM}. 
These advances are welcome and needed in many cases
where next-to-leading order (NLO) calculations are not precise enough,
since higher-order corrections reduce the scale dependence of cross sections
and increase theoretical accuracy.
The search for new physics, such as supersymmetry, in hadron colliders
relies on calculations of both backgrounds and new processes, and this is 
where accuracy becomes important.

Next-to-next-to-leading order calculations are technically difficult
and have been fully completed only for the simplest,
Drell-Yan type \cite{DYNNLO1,DYNNLO2,DYNNLO3,DYNNLO4}, processes.
The corrections can be separated into hard, soft,
and virtual parts, corresponding to contributions from energetic, soft, 
and virtual gluons, respectively.
The soft corrections are an important component of  
the total result both theoretically and numerically. In fact, in certain
kinematical regions, notably threshold, they are dominant.

Recently, a new unified formalism for the calculation
of soft and virtual corrections for arbitrary hard-scattering
processes in hadron-hadron and lepton-hadron collisions,
including QCD, electroweak/Higgs, and supersymmetric processes, was presented 
in Ref. \cite{NKuni}.
It is based on and extends earlier work on threshold resummations
\cite{KS,KOS,LOS,NKrev}.

The calculation of cross sections in hadron-hadron or lepton-hadron
collisions can be written schematically as
\beqa
\sigma=\sum_f \int \left[ \prod_i  dx_i \, \phi_{f/h_i}(x_i,\mu_F)\right]\,
{\hat \sigma}(s,t,u,\mu_F,\mu_R) \, ,
\label{factcs}
\eeqa
where $\sigma$ is the physical cross section,
$\phi_{f/h_i}$ is the distribution function for parton $f$ carrying momentum 
fraction $x_i$ of hadron $h_i$, at a factorization scale $\mu_F$, 
while $\mu_R$ is the renormalization scale.
The parton-level hard-scattering cross section is denoted by 
${\hat \sigma}$, and $s$, $t$, $u$ are standard kinematical invariants
formed from the momentum four-vectors of the particles in the partonic 
reaction. 
In a lepton-hadron collision we have one parton distribution
($i=1$) while in a hadron-hadron collision $i=1,2$.
We note here that $\sigma$ and ${\hat \sigma}$ are not restricted to be
total cross sections; they can represent any relevant differential cross 
section of interest, such as transverse momentum or rapidity distributions.

In general, the partonic cross section $\hat{\sigma}$ includes 
plus distributions ${\cal D}_l(x_{th})$ 
with respect to a kinematical variable $x_{th}$ that measures 
distance from threshold, with $l \le 2n-1$ at $n$-th 
order in $\alpha_s$ beyond the leading order. These are the soft corrections. 
The virtual corrections appear as delta functions, $\delta(x_{th})$.
In single-particle inclusive (1PI) kinematics, 
$x_{th}$ is usually called $s_4$
(or $s_2$), $s_4 \equiv s+t+u-\sum m^2$, and vanishes at threshold. Then
\beq
{\cal D}_l(s_4)\equiv\left[\frac{\ln^l(s_4/M^2)}{s_4}\right]_+ \, ,
\eeq
where $M^2$ is a hard scale relevant to the process at hand, for example
the mass $m$ of a heavy quark, the transverse momentum $p_T$ of a jet, etc.
In pair-invariant-mass (PIM) kinematics, with $Q^2$ the invariant mass 
squared of the produced pair, $x_{th}$ is usually called $1-z$ (or $1-x$), 
with $z=Q^2/s \rightarrow 1$ at threshold.  Then
\beq
{\cal D}_l(z)\equiv\left[\frac{\ln^l(1-z)}{1-z}\right]_+ \, .
\eeq
The logarithms raised to the power $l=2n-1$
are the leading logarithms (LL),
those with $l=2n-2$ are next-to-leading (NLL), 
with $l=2n-3$ are next-to-next-to-leading (NNLL), 
with $l=2n-4$ are next-to-next-to-next-to-leading (NNNLL),
etc. These logarithms can be formally
resummed to all orders in perturbation theory.

In Section 2 I briefly review threshold resummation and present
master formulas for soft and virtual corrections at 
next-to-next-to-leading order for processes in hadron-hadron and 
lepton-hadron collisions. In Section 3 results are given for top
quark pair production at the Tevatron and the LHC, while Section 4 
discusses FCNC top quark production via anomalous couplings
at HERA and the Tevatron. Section 5 discusses $W$ boson hadroproduction
at large transverse momentum at the Tevatron, and Section 6 the production
of a charged Higgs with a top quark at the LHC. 

\mysection{NNLO master formula for soft and virtual corrections}

Resummed cross sections have by now been studied for a variety of processes
\cite{NKrev}.
The exponentiation \cite{KS,KOS,LOS,NKrev}
of soft-gluon corrections arises from the factorization \cite{CSS}
properties of the cross section. The cross section is separated (factorized)
into a short-distance function, $H$, a soft-gluon function, $S$, and parton
distribution functions. The renormalization group evolution of these functions
results in  resummation.

The resummation of threshold logarithms is carried out in moment space.
We define moments of the partonic cross section by
${\hat\sigma}(N)=\int dz \, z^{N-1} {\hat\sigma}(z)$ (PIM) or by \newline  
${\hat\sigma}(N)=\int (ds_4/s) \;  e^{-N s_4/s} {\hat\sigma}(s_4)$ (1PI),
with $N$ the moment variable. Under moments the plus distributions
${\cal D}_l$ give rise to $\ln^{l+1}N$. It is these logarithms of $N$
that exponentiate in moment space to all orders in the strong coupling.
The resummed partonic cross section for any process in hadron-hadron
and lepton-hadron collisions can be written in moment space 
schematically as
\beqa
{\hat{\sigma}}^{res}(N) &=&   
\exp\left[ \sum_i E^{(f_i)}(N_i)+E^{(f_i)}_{\rm scale}(\mu_F,\mu_R)\right] \; 
\exp\left[ \sum_j {E'}^{(f_j)}(N_j)\right] 
\nonumber\\ && \hspace{-10mm} \times \,
{\rm Tr} \left \{H^{f_if_j} \;  \exp \left[\int {d\mu' \over \mu'} \;
{\Gamma}_S^\dagger\right] \; S^{f_if_j} \; 
\exp \left[\int {d\mu' \over \mu'}\; {\Gamma}_S \right] \right\} \, .
\label{resHS}
\eeqa
The sums over $i$ run over incoming partons: in hadron-hadron collisions
$i=1,2$ while in lepton-hadron collisions $i=1$. The sum over $j$ is relevant
only if we have massless partons in the final state at lowest order. 
Clearly the second exponent is then absent for proceses such as Drell-Yan, 
Higgs, and top quark pair production.
Detailed expressions for the exponents are given in Ref. \cite{NKuni}. 
 
The first exponent in Eq. (\ref{resHS}) resums the $N$-dependence of
incoming partons \cite{GS,CT}, including the scale, while  
the second exponent resums the $N$-dependence of any outgoing massless
partons.
The trace appearing in the resummed expression is taken in the space
of color exchanges of the specific process studied.
$H$ is the hard function, independent of any soft-gluon effects, a matrix
in color space.
The evolution of the soft function, $S$, which describes 
noncollinear soft gluon 
emission \cite{KS} and is also a matrix in color space, follows from 
its renormalization group properties and is given in terms of the soft 
anomalous dimension matrix $\Gamma_S$ \cite{KS,KOS,NKrev}. 
In processes with simple color flow, $\Gamma_S$ is a trivial $1\times 1$ 
matrix. For the determination of $\Gamma_S$ in processes with complex 
color flow, an appropriate choice of color basis has to be made.
The process-dependent soft anomalous dimension matrices, evaluated through 
the calculation of eikonal vertex corrections \cite{KS,NK2nl}, 
have by now been presented at one loop for practically all 
partonic processes \cite{NKrev}. 
They can be explicitly calculated for any process using the techniques and 
results in Refs. \cite{KS,NKrev}.
Work is currently being done on 
two-loop calculations of these anomalous dimensions \cite{NK2nl}, 
but we note that the universal component of these
anomalous dimensions for quark-antiquark and gluon-gluon
initiated processes are now known \cite{NKuni}. 

The trace of the product of the hard and soft matrices
at lowest order reproduces the Born cross section for each partonic process.
We use the expansions for the hard and soft matrices:
$H=\alpha_s^{d_{\alpha_s}}H^{(0)}+(\alpha_s^{d_{\alpha_s}+1}/\pi)H^{(1)}
+(\alpha_s^{d_{\alpha_s}+2}/\pi^2)H^{(2)}+\cdots$,
where the constant $d_{\alpha_s}=0,1,2$ if the Born cross section is of order
$\alpha_s^0$, $\alpha_s^1$, $\alpha_s^2$, respectively, and
$S=S^{(0)}+(\alpha_s/\pi)S^{(1)}+(\alpha_s/\pi)^2 S^{(2)}+\cdots$.
The Born term is then given by 
$\sigma^B=\alpha_s^{d_{\alpha_s}} {\rm Tr}[H^{(0)}S^{(0)}]$.

The resummed cross section can be expanded and inverted back
to momentum space to derive master formulas
at any order in the strong coupling for the soft-gluon corrections 
for arbitrary processes in hadron-hadron and lepton-hadron collisions.
We present results in the $\overline {\rm MS}$ scheme. For results
in the DIS scheme see Ref. \cite{NKuni}.

We begin at next-to-leading order. We note that at NLO the 
LL are ${\cal D}_1(x_{th})$ and the NLL are ${\cal D}_0(x_{th})$. 
The virtual corrections multiply $\delta(x_{th})$.
The first-order soft and virtual corrections 
can then be written as \cite{NKuni}
\beq
{\hat{\sigma}}^{(1)} = \sigma^B \frac{\alpha_s(\mu_R^2)}{\pi}
\left\{c_3\, {\cal D}_1(x_{th}) + c_2\,  {\cal D}_0(x_{th}) 
+c_1\,  \delta(x_{th})\right\}+\frac{\alpha_s^{d_{\alpha_s}+1}(\mu_R^2)}{\pi} 
\left[A^c \, {\cal D}_0(x_{th})+T_1^c \, \delta(x_{th})\right] \, ,
\label{NLOcx}
\eeq
where $\sigma^B$ is the Born term, and the LL coefficient is
\beq
c_3=\sum_i 2C_{f_i} -\sum_j C_{f_j}\, ,
\label{c3}
\eeq 
with $C_f=C_F \equiv (N_c^2-1)/(2N_c)$ for quarks and 
$C_f=C_A \equiv N_c$ for gluons,
with $N_c$ the number of colors.
The NLL coefficient $c_2$ is defined by 
\beq
c_2=- \sum_i \left[C_{f_i}
+2 C_{f_i} \, \delta_K \, \ln\left(\frac{-t_i}{M^2}\right)+
C_{f_i} \ln\left(\frac{\mu_F^2}{s}\right)\right]
-\sum_j \left[{B'}_j^{(1)}+C_{f_j}
+C_{f_j} \, \delta_K \, \ln\left(\frac{M^2}{s}\right)\right] \, ,
\label{c2n}
\eeq 
where $M$ is a relevant hard scale and $s$ is the center-of-mass energy 
squared of the incoming partons.
Also,  $\delta_K=1$ for 1PI and 0 for PIM kinematics, 
${B'}_q^{(1)}=3C_F/4$ and ${B'}_g^{(1)}=\beta_0/4$,
where $\beta_0=(11C_A-2n_f)/3$ is the lowest-order beta function,
with $n_f$ the number of light quark flavors. 
Note that the sum over $j$ is only relevant if we have massless partons in 
the final state at lowest order in 1PI kinematics.

The NLL function $A^c$
is process-dependent and depends on the color structure of the 
hard-scattering. It is defined by
\beq
A^c \equiv {\rm Tr} \left(H^{(0)} {\Gamma'}_S^{(1)\,\dagger} S^{(0)}
+H^{(0)} S^{(0)} {\Gamma'}_S^{(1)}\right) \, ,
\label{Ac}
\eeq
where ${\Gamma'}_S^{(1)}$ is the one-loop gauge-independent part of
$\Gamma_S$.
Note that for processes with simple color flow, where $\Gamma_S$ is
a trivial $1\times 1$ matrix, we have
$A^c=\sigma^B \alpha_s^{-d_{\alpha_s}} 2 \, {\rm Re}{\Gamma'}_S^{(1)}$,
where $\rm Re$ denotes the real part.

We also define scale-independent terms
$T_2 \equiv c_2+\sum_i C_{f_i} \ln(\mu_F^2/s)$
and $T_1 \equiv c_1-c_1^{\mu}$,
where the scale-dependent $c_1^{\mu}$ is given explicitly by
\beq
c_1^{\mu}=\sum_i \left[C_{f_i}\, \delta_K \, \ln\left(\frac{-t_i}{M^2}\right) 
-\gamma_i^{(1)}\right]\ln\left(\frac{\mu_F^2}{s}\right)
+d_{\alpha_s} \frac{\beta_0}{4} \ln\left(\frac{\mu_R^2}{s}\right) \, ,
\label{c1mu}
\eeq
with $\gamma_q^{(1)}=3C_F/4$ and $\gamma_g^{(1)}=\beta_0/4$ 
one-loop parton anomalous dimensions for quarks and gluons, respectively.
With respect to the scale-independent virtual $\delta(x_{th})$ terms
$T_1$ and $T_1^c$, we note that they have to be calculated independently 
for each process. 

At next-to-next-to-leading order, the LL are ${\cal D}_3(x_{th})$,
the NLL are ${\cal D}_2(x_{th})$, the NNLL are ${\cal D}_1(x_{th})$,
and the NNNLL are ${\cal D}_0(x_{th})$.
The virtual corrections multiply $\delta(x_{th})$.
The expansion of the resummed cross section, with 
matching to the full NLO soft-plus-virtual result, gives the NNLO
soft and virtual corrections \cite{NKuni}:
\beqa
{\hat{\sigma}}^{(2)}&=&\sigma^B \frac{\alpha_s^2(\mu_R^2)}{\pi^2}
\left\{\frac{1}{2} c_3^2 \, {\cal D}_3(x_{th})
+\left[\frac{3}{2} c_3 \, c_2 
- \frac{\beta_0}{4} c_3
+\sum_j C_{f_j} \frac{\beta_0}{8}\right] {\cal D}_2(x_{th})\right.
\nonumber \\ && \hspace{-5mm}
{}+\left[c_3 \, c_1 +c_2^2
-\zeta_2 \, c_3^2 -\frac{\beta_0}{2} \, T_2 
+\frac{\beta_0}{4} c_3  \ln\left(\frac{\mu_R^2}{s}\right)
+\sum_i C_{f_i} \, K \right.
\nonumber \\ && \hspace{-5mm} \quad \quad \left.
{}+\sum_j C_{f_j} \left(-\frac{K}{2} 
+\frac{\beta_0}{4} \, \delta_K \,  \ln\left(\frac{M^2}{s}\right)\right)
-\sum_j\frac{\beta_0}{4} {B'}_j^{(1)} \right]
{\cal D}_1(x_{th})
\nonumber \\ && \hspace{-5mm} 
{}+\left[c_2 \, c_1 -\zeta_2 \, c_2 \, c_3+\zeta_3 \, c_3^2 
-\frac{\beta_0}{2} T_1
+\frac{\beta_0}{4}\, c_2 \ln\left(\frac{\mu_R^2}{s}\right) 
-{\cal G}_{ij}^{(2)}
\right. 
\nonumber \\ && \hspace{-5mm} \quad \quad
{}+\sum_i C_{f_i} \left(\frac{\beta_0}{8} 
\ln^2\left(\frac{\mu_F^2}{s}\right)
-\frac{K}{2}\ln\left(\frac{\mu_F^2}{s}\right)
-K \, \delta_K \, \ln\left(\frac{-t_i}{M^2}\right)\right)
\nonumber \\ && \hspace{-5mm} \quad \quad \left.\left.
{}+\sum_j C_{f_j} \, \delta_K \,   \left(\frac{\beta_0}{8}
\ln^2\left(\frac{M^2}{s}\right)
-\frac{K}{2}\ln\left(\frac{M^2}{s}\right)\right)
-\sum_j \frac{\beta_0}{4}
{B'}_j^{(1)} \, \delta_K \,  \ln\left(\frac{M^2}{s}\right) \right]
{\cal D}_0(x_{th})\right\}
\nonumber \\ &&  \hspace{-10mm}
{}+\frac{\alpha_s^{d_{\alpha_s}+2}(\mu_R^2)}{\pi^2} 
\left\{\frac{3}{2} c_3 \, A^c\, {\cal D}_2(x_{th})
+\left[\left(2c_2-\frac{\beta_0}{2}\right)A^c+c_3 T_1^c
+F^c\right] {\cal D}_1(x_{th})
\right.
\nonumber \\ &&  \left.
{}+\left[\left(c_1-\zeta_2 c_3+\frac{\beta_0}{4}
\ln\left(\frac{\mu_R^2}{s}\right)\right)A^c+\left(c_2-\frac{\beta_0}{2}\right)
T_1^c+F^c \, \delta_K \,\ln\left(\frac{M^2}{s}\right)+G^c\right] 
{\cal D}_0(x_{th})\right\}
\nonumber \\ && \hspace{-10mm}
{}+R^{(2)}_{\delta}  \delta(x_{th}) \, ,
\label{NNLOsoft}
\eeqa
where $\zeta_2=\pi^2/6$ and $\zeta_3=1.2020569\cdots$, 
$G_{ij}^{(2)}$ and $G^c$ are process-dependent two-loop functions
\cite{NKuni,NK2nl}, 
$K=C_A (67/18-\pi^2/6)-5n_f/9$ \cite{KT}, and 
\beq
F^c={\rm Tr} \left[H^{(0)} \left({\Gamma'}_S^{(1)\,\dagger}\right)^2 S^{(0)}
+H^{(0)} S^{(0)} \left({\Gamma'}_S^{(1)}\right)^2
+2 H^{(0)} {\Gamma'}_S^{(1)\,\dagger} S^{(0)} {\Gamma'}_S^{(1)} \right] \, .
\label{Fterm}
\eeq

We cannot provide a universal equation for the full virtual terms 
$R^{(2)}_{\delta}$, although several  terms have been given in
Ref. \cite{NKuni}. However, the $\delta(x_{th})$ terms that involve 
the factorization and renormalization scales can be given fully and explicitly.
Let $C^{\mu}_{t_i}\equiv \sum_i \left[C_{f_i} \, \delta_K \, 
\ln\left(-t_i/M^2\right)-\gamma_i^{(1)}\right]$.
Then those scale-dependent terms are given by
\beqa
R^{\mu\, (2)}_{\delta}&=&\sigma^B \frac{\alpha_s^2(\mu_R^2)}{\pi^2}
\left\{\ln^2\left(\frac{\mu_F^2}{M^2}\right)
\left[\frac{1}{2} (C^{\mu}_{t_i})^2
-\frac{\zeta_2}{2}\left(\sum_i C_{f_i}\right)^2
- \frac{\beta_0}{8} C^{\mu}_{t_i}\right]\right.
\nonumber \\ &&
{}+\ln\left(\frac{\mu_F^2}{M^2}\right)\ln\left(\frac{\mu_R^2}{M^2}\right)
\frac{\beta_0}{4}(d_{\alpha_s}+1) C^{\mu}_{t_i} 
+\ln^2\left(\frac{\mu_R^2}{M^2}\right)
\frac{\beta_0^2}{32}(d_{\alpha_s}^2+d_{\alpha_s})
\nonumber \\ &&
{}+\ln\left(\frac{\mu_F^2}{M^2}\right)
\left[(C^{\mu}_{t_i})^2 \, \delta_K \, \ln\left(\frac{M^2}{s}\right)
+C^{\mu}_{t_i} \left(T_1+d_{\alpha_s}\frac{\beta_0}{4}
\, \delta_K \, \ln\left(\frac{M^2}{s}\right)\right)
-\sum_i {\gamma'}_{i/i}^{(2)}\right.
\nonumber \\ && \left.
{}+(\sum_i C_{f_i}) \left(\zeta_2 \left(T_2-\sum_i C_{f_i}
\, \delta_K \, \ln\left(\frac{M^2}{s}\right)\right)
-\zeta_3 c_3
+\frac{K}{2} \, \delta_K \, \ln\left(\frac{-t_i}{M^2}\right)\right)\right]
\nonumber \\ && \left.
{}+\ln\left(\frac{\mu_R^2}{M^2}\right)
\left[\frac{\beta_0}{4}(d_{\alpha_s}+1) 
\left(C^{\mu}_{t_i} \, \delta_K \, \ln\left(\frac{M^2}{s}\right)
+d_{\alpha_s}\frac{\beta_0}{4} \, \delta_K \, 
\ln\left(\frac{M^2}{s}\right)+T_1\right)
+\beta_1  \frac{d_{\alpha_s}}{16}\right]\right\}
\nonumber \\ && \hspace{-5mm}
{}+\frac{\alpha_s^{d_{\alpha_s}+2}}{\pi^2}
\left\{\left(C^{\mu}_{t_i} T_1^c+\zeta_2 \sum_i C_{f_i} A^c \right) 
\ln\left(\frac{\mu_F^2}{M^2}\right) 
+(d_{\alpha_s}+1) \frac{\beta_0}{4}\ln\left(\frac{\mu_R^2}{M^2}\right)
T_1^c\right\} \, ,
\label{NNLOvs}
\eeqa
where $\beta_1=34 C_A^2/3-2n_f(C_F+5C_A/3)$ is the NLO beta function, and
\beq
{\gamma'}_{q/q}^{(2)}=C_F^2\left(\frac{3}{32}-\frac{3}{4}\zeta_2
+\frac{3}{2}\zeta_3\right)
+C_F C_A\left(-\frac{3}{4}\zeta_3+\frac{11}{12}\zeta_2+\frac{17}{96}\right)
+n_f C_F \left(-\frac{\zeta_2}{6}-\frac{1}{48}\right)\, ,
\eeq
and
\beq
{\gamma'}_{g/g}^{(2)}=C_A^2\left(\frac{2}{3}+\frac{3}{4}\zeta_3\right)
-n_f\left(\frac{C_F}{8}+\frac{C_A}{6}\right) 
\eeq
are two-loop anomalous dimensions for quarks and gluons \cite{NKuni}.

As shown in Ref. \cite{NKuni} further virtual corrections
can also be deduced, including $\zeta$ terms that arise from the
inversion from moment to momentum space. For further details
we refer the reader to Ref. \cite{NKuni}.
Eqs. (\ref{NNLOsoft}) and (\ref{NNLOvs}) constitute master formulas
for the calculation of NNLO soft-gluon corrections and virtual corrections
involving the scale for arbitrary processes in hadron-hadron and 
lepton-hadron colliders. They will be applied in the following sections
to various processes of phenomenological interest.

\mysection{Top quark production}

As the first application of the unified formalism we consider
top quark production, recently studied in Ref. \cite{KV}.
The top quark is currently being re-observed at Run II of the Tevatron
\cite{Shab,D0,Azzi}, 
and it is expected that improvements in the measurements of the values 
of the top mass and production cross section will be achieved.

The cross section for $t {\bar t}$ production can be calculated
in both 1PI and PIM kinematics. 
When not all terms are known in the theoretical expression
for the cross section at NNLO, there is some difference
between single-particle-inclusive (1PI) and pair-invariant-mass
(PIM) kinematics. When NNNLL terms are included \cite{KV},
the kinematics dependence of the cross section vanishes
near threshold and is reduced away from it relative to NNLL 
accuracy \cite{NKtop,KLMV}. 
The factorization and renormalization
scale dependence of the cross section is also greatly reduced.
Thus top production is brought under theoretical control.

We study the partonic processes $ij \rightarrow t {\overline t}$ 
with $ij = q {\bar q}$ and $gg$.  
In 1PI kinematics, a single top quark $t$ is identified,
$i(p_a) + j(p_b) \longrightarrow t(p_1) + X[{\overline t}](p_2)$,
where $X[{\overline t}]$ is the remaining final state that contains
the ${\overline t}$.
We define the kinematical invariants 
$s=(p_a+p_b)^2$, $t_1=(p_b-p_1)^2-m^2$, $u_1=(p_a-p_1)^2-m^2$, with $m$ the 
top quark mass, and $s_4=s+t_1+u_1$. At threshold, $s_4 \rightarrow 0$,
and the soft corrections appear as $[\ln^l(s_4/m^2)/s_4]_+$.
In PIM kinematics, we have instead
$i(p_a) + j(p_b) \longrightarrow t{\overline t}(p) + X(k)$.
At partonic threshold, $s=M^2$, with $M^2$ the pair-mass squared.
The soft corrections appear as $[\ln^l(1-z)/(1-z)]_+$,
with $z=M^2/s \rightarrow 1$ at threshold.
In either kinematics we calculate the NNLO soft corrections to the double
differential cross section and add them to the exact NLO \cite{NLOqq,NLOgg} 
result.

The partonic scaling functions for top production
were studied in detail in Ref. \cite{KV}.
Note that to NNLL, the two kinematics choices give rather different results,
even near threshold.  When the NNNLL contribution is added, both the 1PI and
PIM results are reduced relative to NNLL and the agreement 
between the two kinematics is much improved.  Adding
virtual $\zeta$ terms resulting from inversion of the resummed cross section
from moment to momentum space improves the agreement 
between the 1PI and PIM kinematics further away from threshold. 
We note here that the effect of the virtual $\zeta$ terms is numerically 
small, as is also the case for the hadronic results for both channels. 
This small effect is in agreement with the arguments in
Ref.~\cite{NKtop} concerning resummation prescriptions,
where it was shown that when subleading terms from inversion are
calculated exactly they do not have an unwarranted large effect on the
numerical results.

\begin{figure}[htpb]
\epsfxsize=0.44\textwidth\epsffile{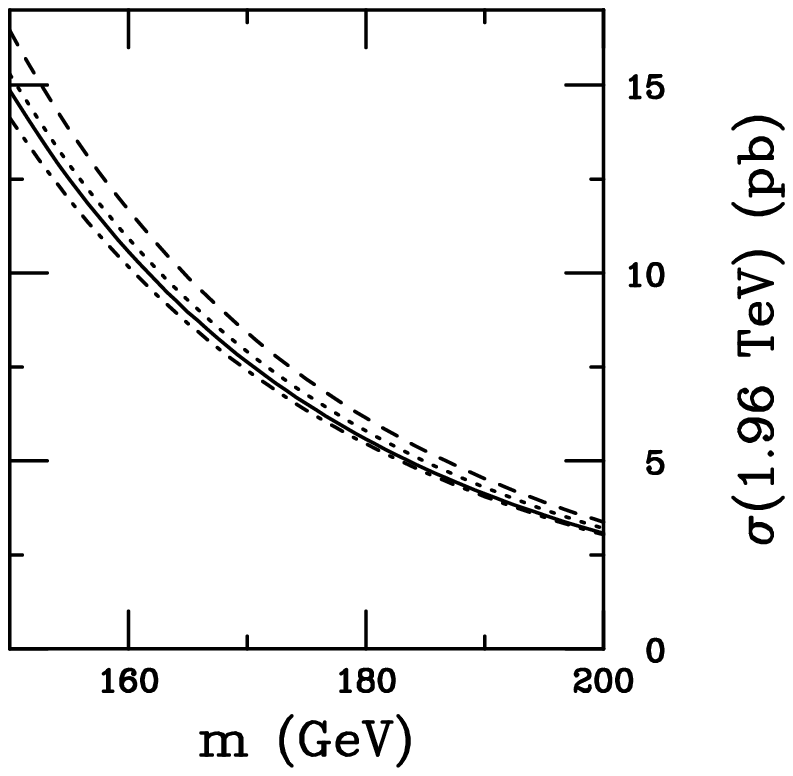}
\epsfxsize=0.56\textwidth\epsffile{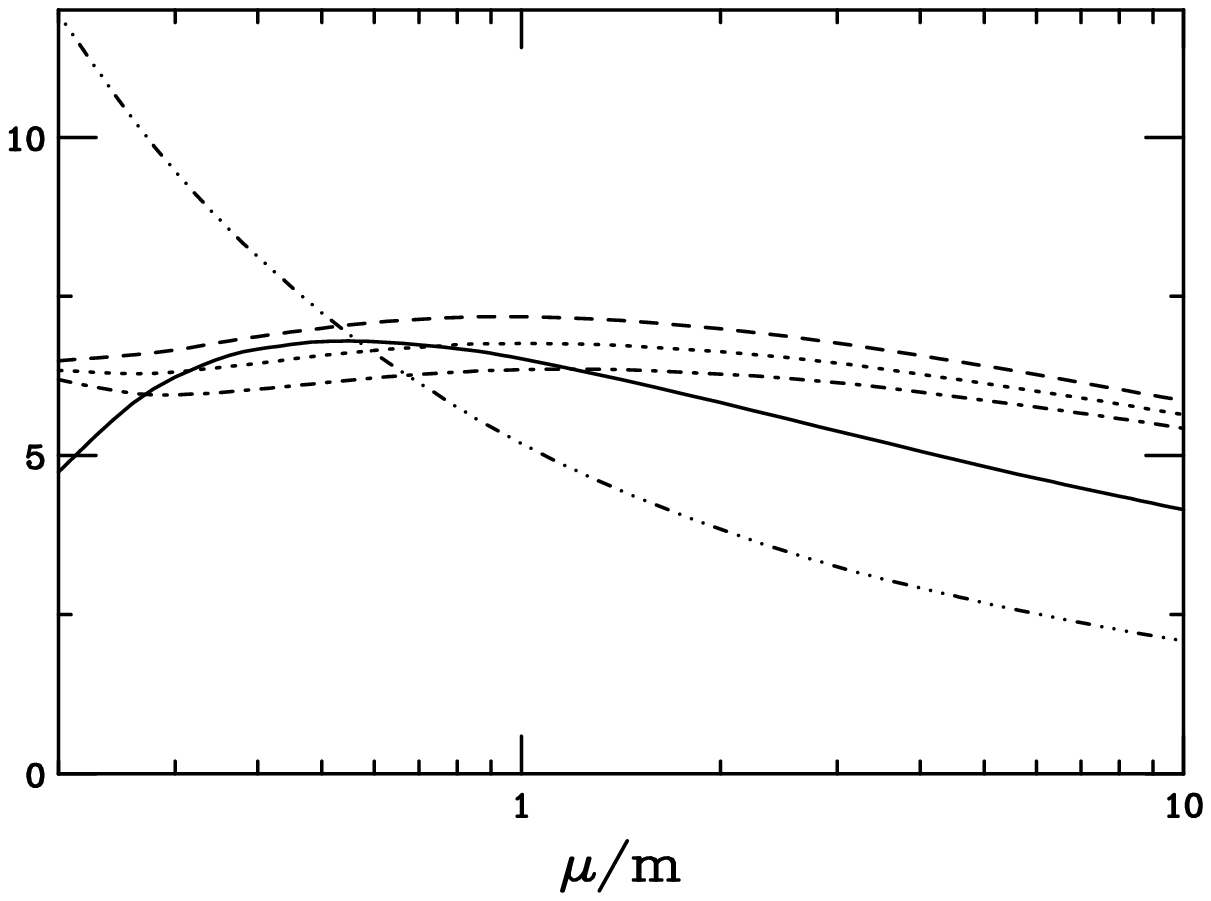}
\caption[]{The $t \overline t$ total cross section in $p \overline p$
collisions at $\sqrt{S} = 1.96$ TeV is shown as a function of $m$ for
$\mu = m$ (left frame), and $\mu/m$ for $m=175$ GeV (right frame).  
The LO (dot-dot-dot-dashed), NLO (solid), and NNLO-NNNLL+$\zeta$ 1PI (dashed), 
PIM (dot-dashed) and average (dotted) results are plotted.}
\label{ttbarcs}
\end{figure}

In Fig. \ref{ttbarcs} we present the NLO and 
soft NNLO $t \overline t$ cross sections 
in $p{\bar p}$ collisions at the Tevatron with $\sqrt{S} = 1.96$ TeV
as functions of mass and scale \cite{KV}.  We use the MRST2002 NNLO parton
densities \cite{MRST2002} (results with the CTEQ6M densities 
\cite{CTEQ6M} are similar). 
The NNLO results include the soft NNNLL and virtual $\zeta$ terms,
and thus denoted NNLO-NNNLL+$\zeta$, in 1PI and PIM kinematics.  
We also show the average of the two kinematics results
which may perhaps be closer to the full NNLO result.  
In the left frame we plot the cross section as a function of top quark mass
for $\mu\equiv\mu_F=\mu_R=m$.
In the right frame we plot the scale dependence of the cross section
over two orders of magnitude in $\mu/m$ with $m=175$ GeV.
The NLO cross section has a milder dependence on scale than 
the leading order (LO) result.
The NNLO cross section exhibits even less dependence on $\mu/m$, 
approaching the independence of scale corresponding to a truly physical 
cross section. The change in the NNLO cross section in the
range $m/2 < \mu < 2m$, normally displayed as a measure of uncertainty
from scale variation, is less than 3\%.

\begin{figure}[htpb] 
\epsfxsize=0.5\textwidth\epsffile{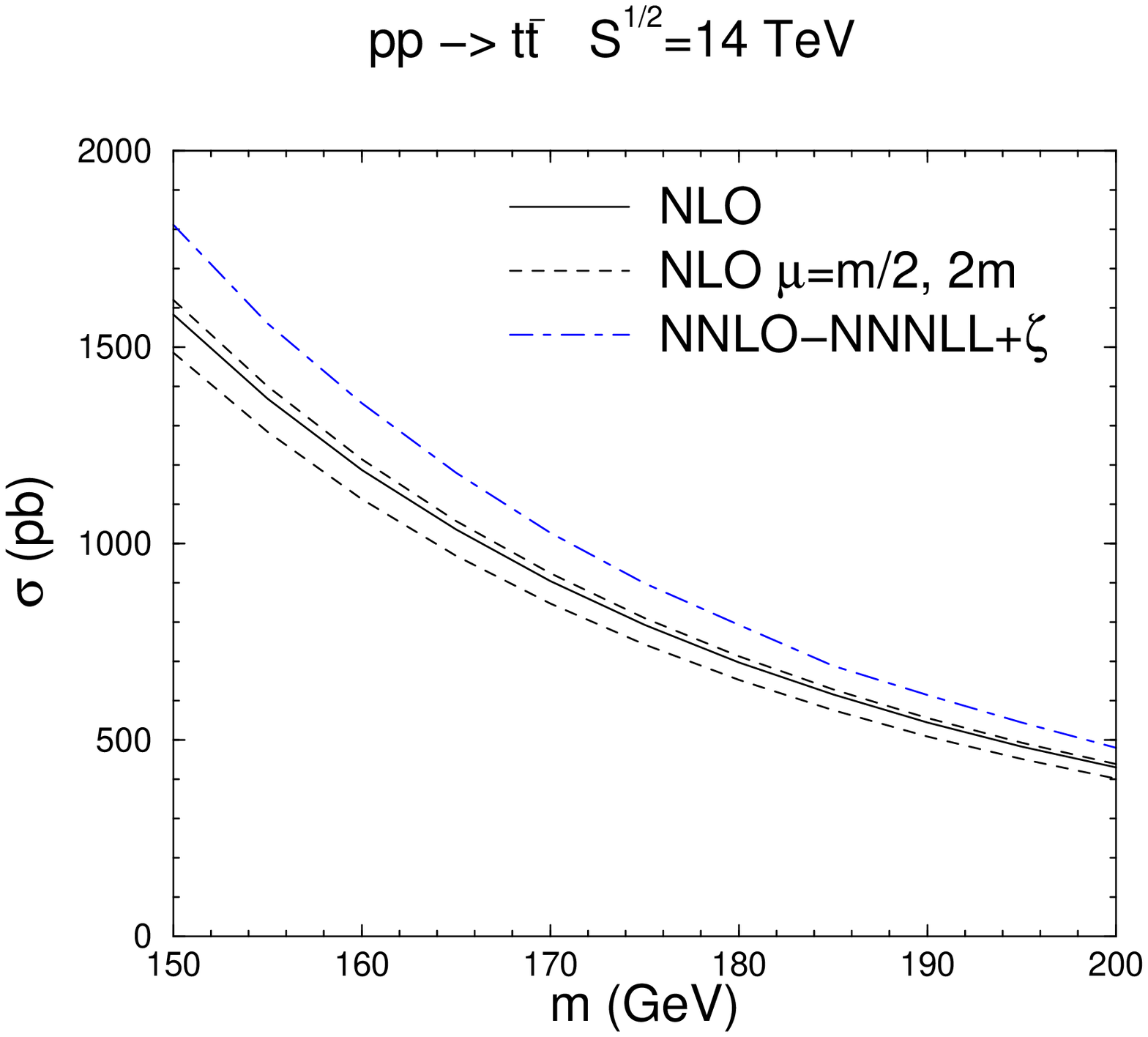}
\epsfxsize=0.5\textwidth\epsffile{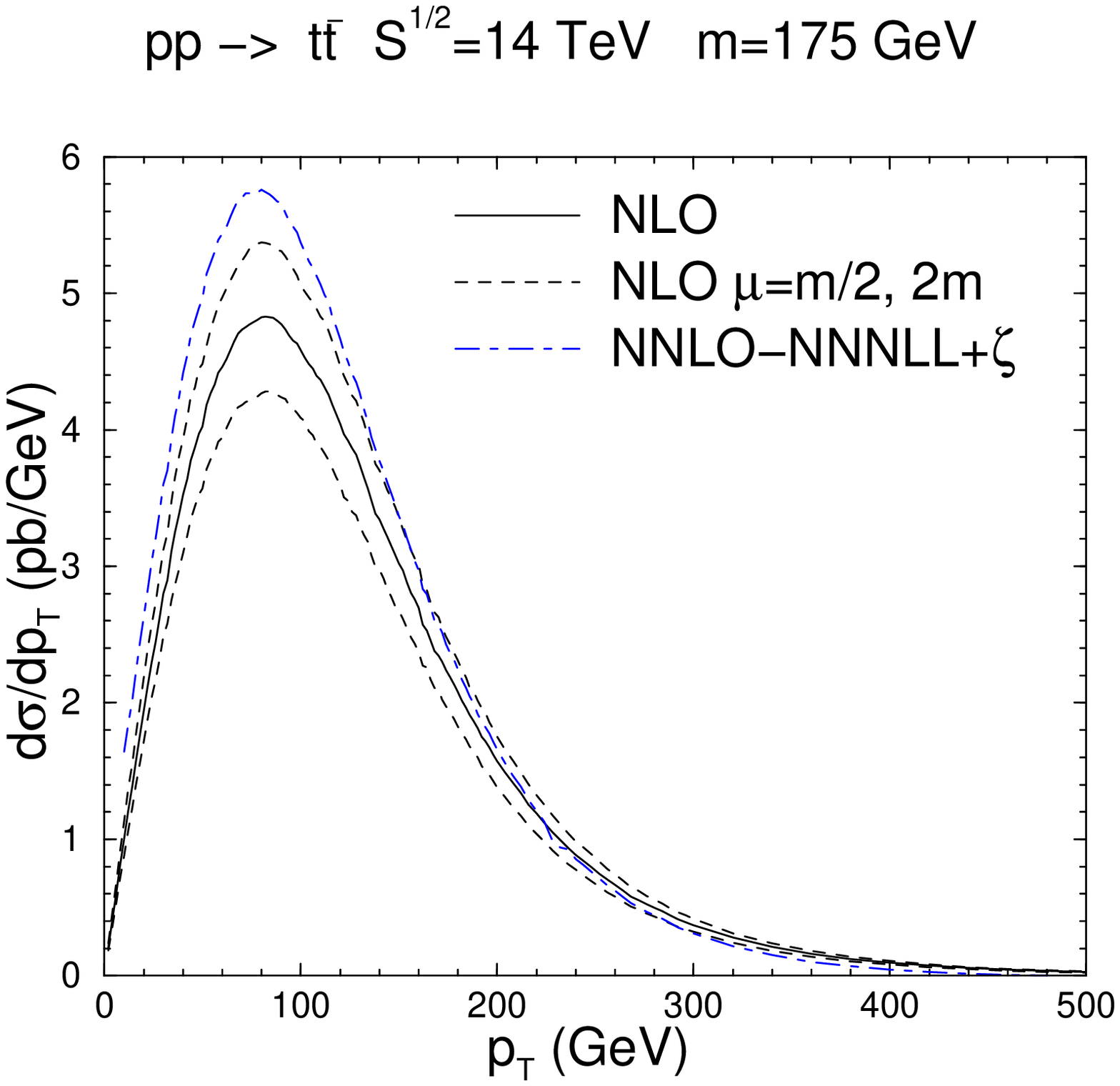}
\caption[]{$t {\bar t}$ production in $pp$ collisions at the LHC
with $\sqrt{S} = 14$ TeV.
Left: The mass dependence of the total cross section. 
Right: The top quark $p_T$ distribution with $m = 175$ GeV. The
NLO (solid), and NNLO-NNNLL+$\zeta$ 1PI (dot-dashed) results are 
shown for $\mu=m$. The NLO results are also shown for $\mu=m/2$ (upper dashed)
and $2m$ (lower dashed).} 
\label{toplhc}
\end{figure}

In Fig. \ref{toplhc} we present results in 1PI kinematics 
for top production in $pp$ collisions
at the LHC with $\sqrt{S}=14$ TeV. 
We note that PIM results are unreliable for processes where the gluon 
contribution is dominant, such as top production at the LHC \cite{KV}
and bottom and charm production at fixed-target experiments \cite{KVbc}.
The left frame shows the total cross section
versus the top mass.
In the right frame the top quark transverse 
momentum distribution is plotted. 
At NNLO we observe an enhancement of the NLO $p_T$ distribution
with no significant change in shape. 

\mysection{FCNC top quark production via anomalous $tqV$ couplings}

Recently NNLO threshold corrections were calculated for top quark production 
via anomalous $tqV$ couplings in flavor-changing 
neutral-current (FCNC) processes at the Tevatron and HERA colliders
\cite{ABNK,NKAB}.

The study of anomalous couplings involving the top quark
is well motivated in various models involving new physics.
The effective Lagrangian involving such  couplings of a $t,q$ pair 
to massless bosons is the following:
\begin{equation}
\Delta {\cal L}^{eff} =    \frac{1}{ \Lambda } \,
\kappa_{tqV} \, e \, \bar t \, \sigma_{\mu\nu} \, q \, F^{\mu\nu}_V + h.c.,
\label{fcnc-eq}
\end{equation}
where $\kappa_{tqV}$ is the anomalous FCNC coupling, with 
$q$ a $u$- or $c$-quark and $V$ a photon or $Z$-boson;
$F^{\mu\nu}_V$  are the usual photon/Z-boson field tensors;
$\sigma_{\mu \nu}=(i/2)(\gamma_{\mu}\gamma_{\nu}
-\gamma_{\nu}\gamma_{\mu})$ with $\gamma_{\mu}$ the Dirac matrices;
$e$ is the electron charge; and $\Lambda$ is an effective scale which 
we will take to be equal to the top quark mass, denoted by $m$.

Current  colliders such as the Tevatron 
and HERA have a good potential to search for FCNC 
interactions in the top-quark sector. In order to establish accurate limits
on anomalous couplings, experiments need accurate predictions of cross sections
for FCNC processes. It was shown in Refs. \cite{ABNK,NKAB} that at HERA 
uncertainties for the tree-level FCNC cross section 
$ep \rightarrow et$ can be very big, since
the cross section can vary by a factor of two 
due to the choice of the factorization and renormalization scale. 
Similar conclusions were drawn in Ref. \cite{NKAB}    
for the Born-level cross sections for the FCNC processes
$p{\bar p} \rightarrow tZ$, $p{\bar p} \rightarrow t\gamma$,
and  $p{\bar p} \rightarrow tt$ at the Tevatron.
This fact unavoidably requires the calculation of
higher-order corrections for the stabilization of the FCNC cross section.

\begin{figure}[htb] 
\epsfxsize=0.5\textwidth\epsffile{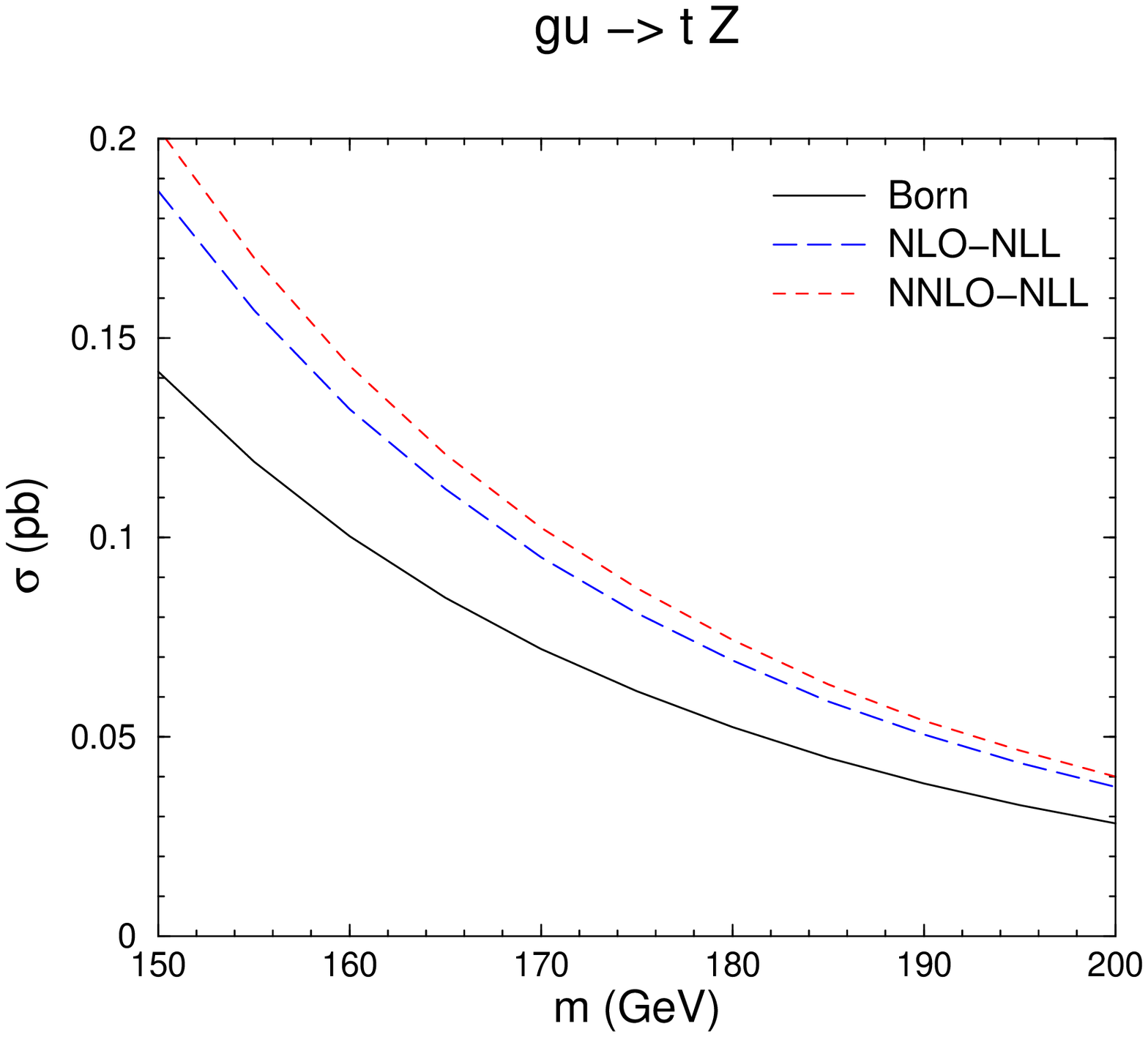}
\epsfxsize=0.5\textwidth\epsffile{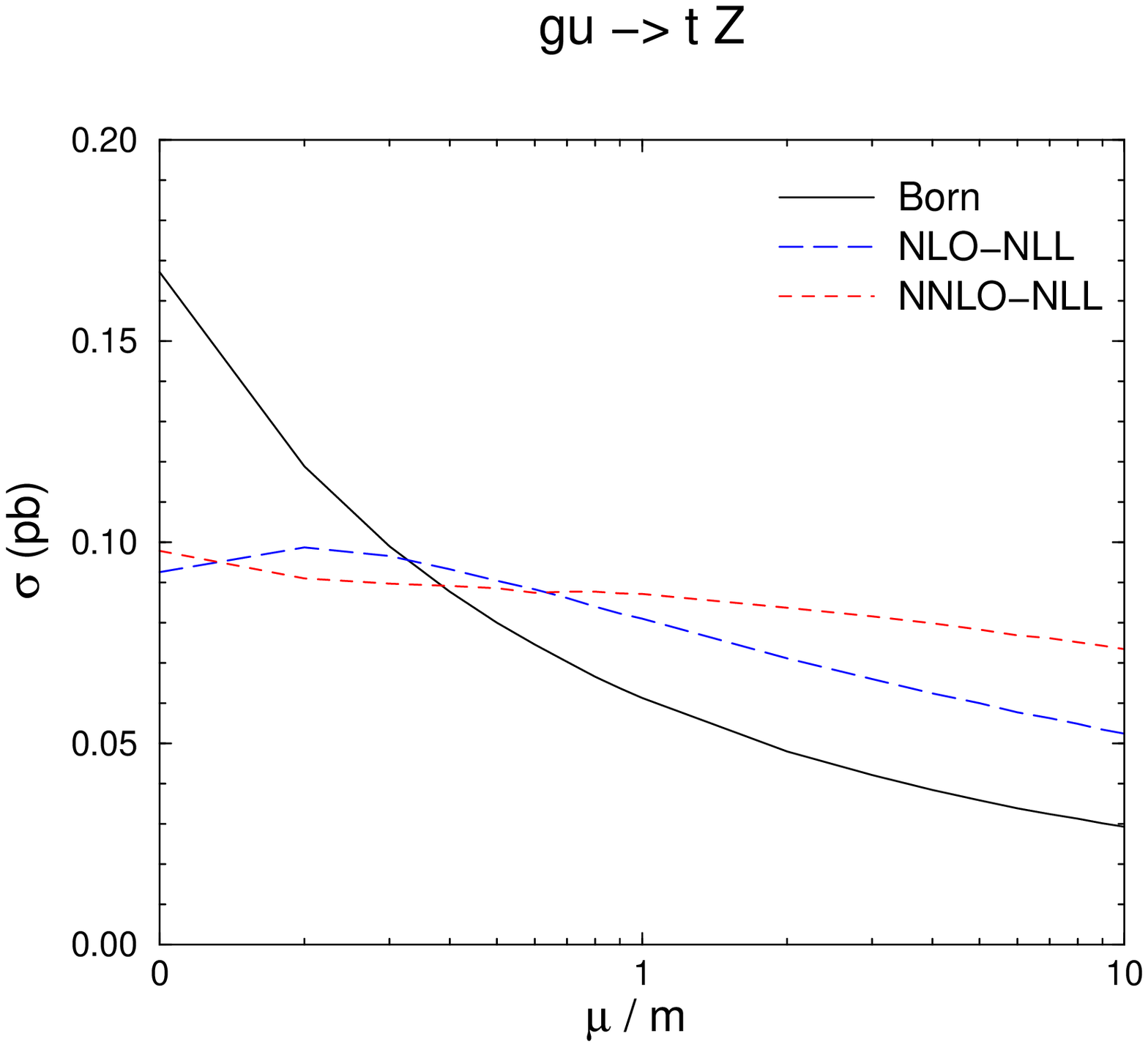}
\caption[]{The Born, NLO-NLL, and NNLO-NLL cross sections
for $gu\rightarrow tZ$ in $p \bar p$ collisions
with $\sqrt{S}=1.96$ TeV and $\kappa_{tuZ}=0.1$.
Left:  cross section versus top-quark mass at scale $\mu=m$.
Right: cross section versus $\mu/m$ with $m=175$ GeV.
}
\label{gutZ} 
\end{figure}

As was shown in Refs. \cite{ABNK,NKAB} the inclusion of NLO and NNLO
threshold corrections indeed greatly stabilizes the cross sections
for the aforementioned FCNC processes. 
For $ep \rightarrow et$ the uncertainty due to scale variation
drops to around 10 percent.This is of particular importance for HERA
studies of FCNC processes \cite{HERA1,HERA2,HERA3,HERA4,HERA5}.

Similarly, for the several possible FCNC processes at the Tevatron
the theoretical predictions are stabilized with inclusion
of threshold corrections. As an example, we consider
the FCNC process $p{\bar p} \rightarrow tZ$ \cite{AAS,TTY}
at the Tevatron, which at the partonic level proceeds mainly through 
$gu \rightarrow tZ$ via an anomalous
$\kappa_{tuZ}$ coupling (the contribution from $gc \rightarrow tZ$ is much
smaller). The NLO and NNLO threshold corrections at NLL accuracy were
calculated in Ref. \cite{NKAB}.
The left frame of Fig. \ref{gutZ} shows the total cross section versus top 
quark mass for scale $\mu=m$. We see that the NLO-NLL threshold corrections
provide an important enhancement of the Born result. The NNLO-NLL
corrections provide an additional increase of the cross section.
The right frame plots the scale dependence of the cross section
over two orders of magnitude: $0.1 < \mu/m < 10$. We note the significant
stabilization of the cross section when the NLO and NNLO threshold 
corrections are included.

\mysection{$W$-boson hadroproduction at large transverse momentum}

The production of $W$ bosons  in hadron colliders  
is a process of relevance in testing the Standard Model, 
calculating backgrounds to new physics such as
associated Higgs boson production, and luminosity monitoring.

The calculation of the complete NLO
cross section for $W$ hadroproduction at large transverse momentum was 
presented in Refs. \cite{PAMR,GPW}.
The NLO corrections enhance the 
differential distributions in transverse momentum $Q_T$ of the $W$ boson.
The $Q_T$ distribution falls rapidly with increasing $Q_T$, spanning six
orders of magnitude in the 10 GeV $< Q_T <$ 190 GeV region 
at the Tevatron. 

$W$-boson production at high transverse momentum receives important 
corrections from the emission of soft gluons from the partons in the process.
Soft-gluon resummation and NNLO-NNLL corrections for this process
were studied in Ref. \cite{NKVD}. More recently the NNLO-NNNLL corrections
were studied in Ref. \cite{NKASV}.
These threshold corrections further enhance the cross section and
reduce the scale dependence \cite{NKASV}.

For the hadronic production of a high-$Q_T$ $W$ boson, with mass $m_W$,
the lowest-order partonic subprocesses are  
$q(p_a)+g(p_b) \longrightarrow W(Q) + q(p_c)$ and
$q(p_a)+{\bar q}(p_b) \longrightarrow W(Q) + g(p_c)$. 
The partonic kinematical invariants in the process are 
$s=(p_a+p_b)^2$, $t=(p_a-Q)^2$, $u=(p_b-Q)^2$, which satisfy 
$s_2 \equiv s + t + u - Q^2=0$ at partonic threshold.
Note that $s_2=(p_a+p_b-Q)^2$ is the invariant mass of the system recoiling 
against the $W$ boson.
The partonic cross section $\hat{\sigma}$ 
includes distributions with respect to $s_2$ of the type
$[\ln^{l}(s_2/Q_T^2)/s_2]_+$.

\begin{figure}[htpb] 
\epsfxsize=0.5\textwidth\epsffile{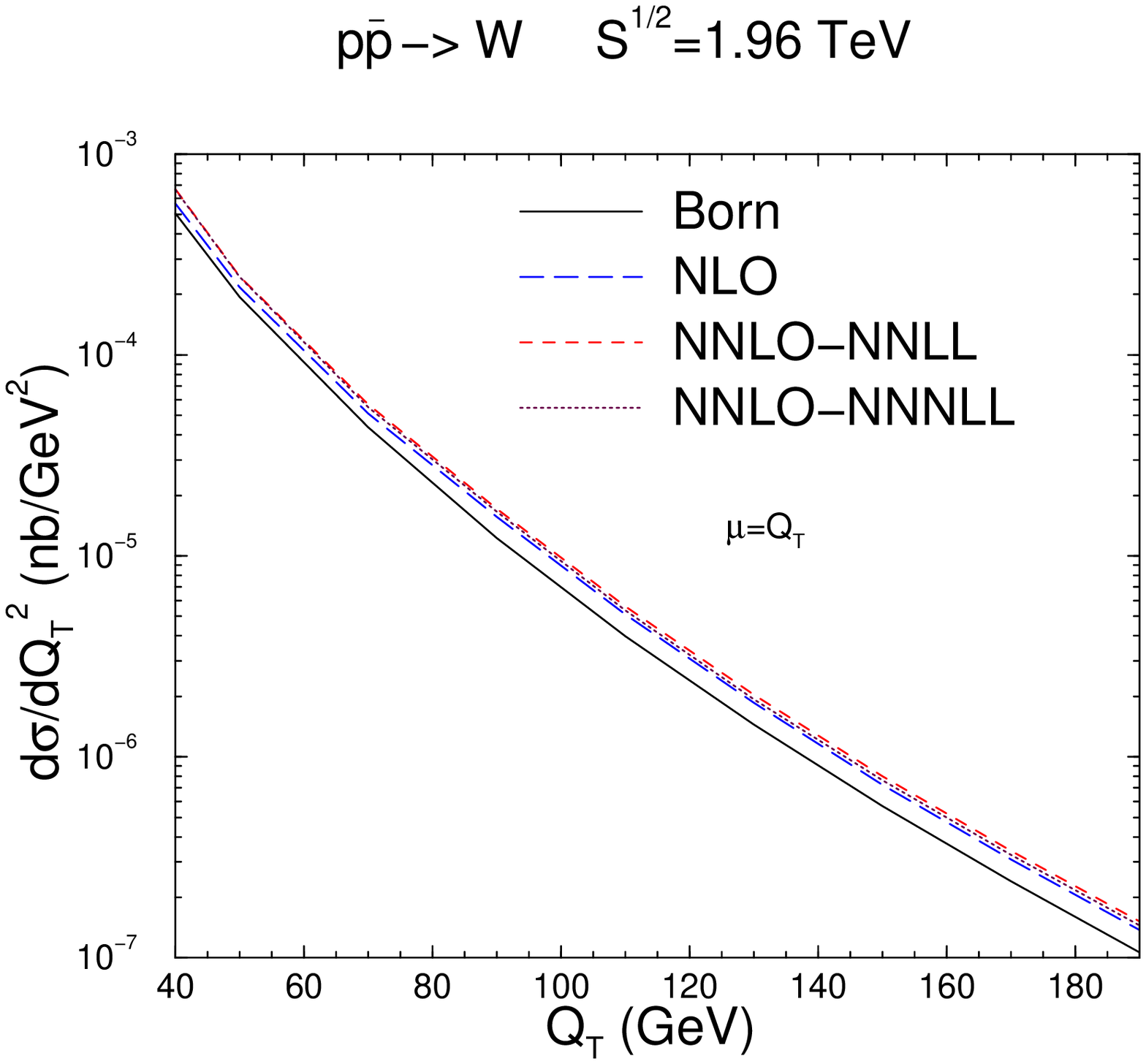}
\epsfxsize=0.5\textwidth\epsffile{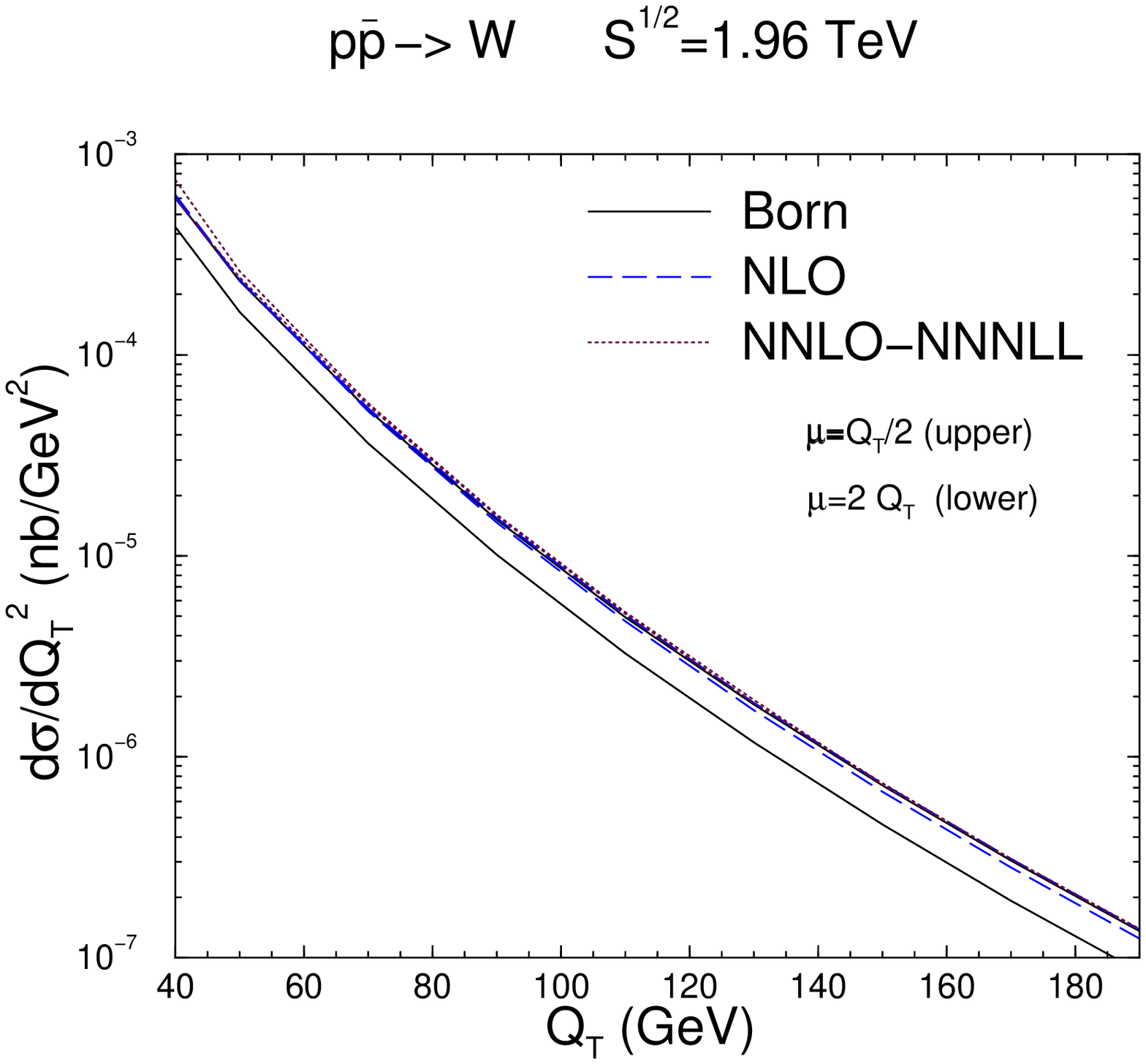}
\caption[]{The differential cross section,
$d\sigma/dQ_T^2$, for $W$ hadroproduction in $p \bar p$ collisions
at the Tevatron with $\sqrt{S}=1.96$ TeV.
Left frame: We plot the Born, NLO, NNLO-NNLL, and NNLO-NNNLL results
with $\mu\equiv\mu_F=\mu_R=Q_T$.
Right frame: We plot the Born, NLO, and NNLO-NNNLL results
with $\mu=Q_T/2$ and $2Q_T$.}
\label{WQTplots}
\end{figure}

In Fig. \ref{WQTplots} we plot the the transverse momentum distribution,
$d\sigma/dQ_T^2$, for $W$ hadroproduction at the Tevatron 
Run II with $\sqrt{S}=1.96$ TeV.
We use the MRST2002 NNLO parton densities \cite{MRST2002}.
In the left frame,
we set the factorization scale $\mu_F$ and the renormalization scale
$\mu_R$ equal to $Q_T$.
We focus on the large $Q_T$ region where the soft-gluon approximation holds
well and the corrections are important.
We see that the NLO corrections provide a significant enhancement of the Born
cross section. The NNLO-NNLL corrections provide a further
modest enhancement of the $Q_T$ distribution. 
If we increase the accuracy by including the NNNLL contributions,
which are negative,
then we find that the NNLO-NNNLL cross section lies between
the NLO and NNLO-NNLL results.
In the right frame, we vary the scale by a factor of two.
We see that the NNLO-NNNLL distribution is more stable under
scale variation than the NLO result, which in turn is much more stable than LO.
In fact the two NNLO-NNNLL curves and the $\mu=Q_T/2$ Born and NLO curves
all lie on top of each other.

The $K$-factors were studied in Ref. \cite{NKASV} where it was shown that they
are moderate and nearly constant over the high-$Q_T$ range  even though
the distributions themselves span a few orders of magnitude.
It was also shown that in the high-$Q_T$ region
the soft-gluon approximation holds very well, as the NLO-NLL cross
section is almost identical to the full NLO result. 
The scale dependence of the differential cross section
was also studied over two orders of magnitude, $0.1 < \mu/Q_T < 10$, 
at fixed $Q_T$. 
There is good stabilization of the cross section 
over the entire range in $\mu/Q_T$ when the NLO corrections are
included, and further improvement when the NNLO-NNNLL corrections
are added.

Finally, we note that there are related NNLO threshold results for direct 
photon production \cite{NKJO}, a process for which the partonic 
subprocesses are essentially the same as for $W$ hadroproduction.

\mysection{Charged Higgs production via the process $bg \longrightarrow t H^-$}

The discovery of the Higgs boson(s) is one of the main aims of the current
run at the Tevatron and the future program at the LHC.
The Minimal Supersymmetric Standard Model (MSSM) 
introduces charged Higgs bosons 
in addition to the Standard Model neutral Higgs boson.
The charged Higgs would be an important signal of new physics
beyond the Standard Model. 

An important partonic channel for charged Higgs discovery
at the LHC is associated production with
a top quark via bottom-gluon fusion, 
$b(p_b)+g(p_g) \rightarrow t(p_t)+H^-(p_H)$,
for which we define the kinematical invariants $s=(p_b+p_g)^2$,
$t=(p_b-p_t)^2$, $u=(p_g-p_t)^2$, and $s_4=s+t+u-m_t^2-m_H^2$,
where $m_H$ is the charged Higgs mass, $m_t$ is the top quark mass,
and we ignore the bottom quark mass $m_b$. 
Threshold corrections appear as $[\ln(s_4/m_H^2)/s_4]_+$.
The next-to-leading order (NLO) corrections to this process
were calculated in Refs. \cite{Zhu,Plehn}.
Here, I discuss soft-gluon corrections to charged Higgs
production, which are expected to be important near threshold, 
the kinematical region where the charged Higgs may be discovered 
at the LHC. 

\begin{figure}[htpb] 
\epsfxsize=0.5\textwidth\epsffile{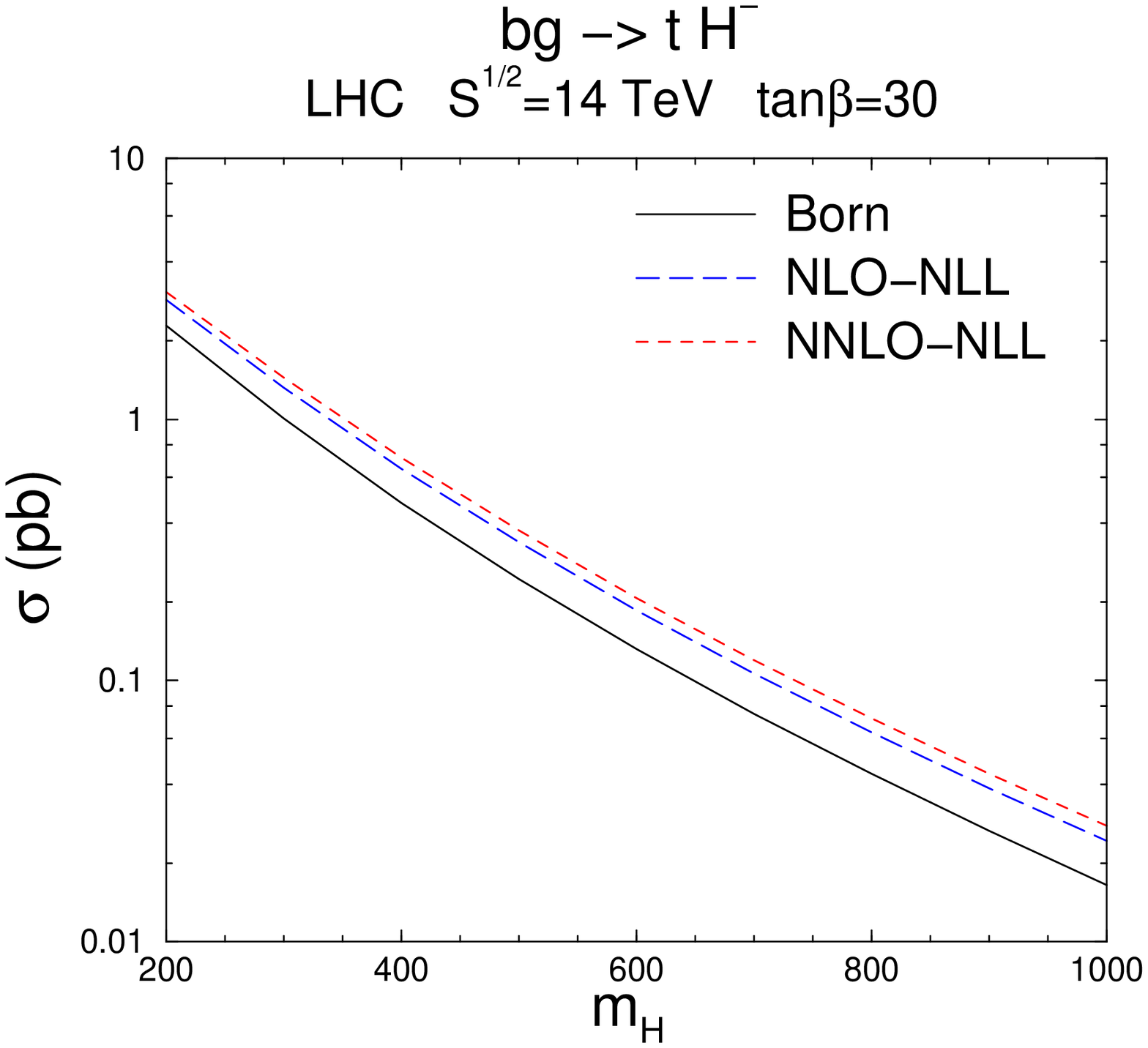}
\epsfxsize=0.5\textwidth\epsffile{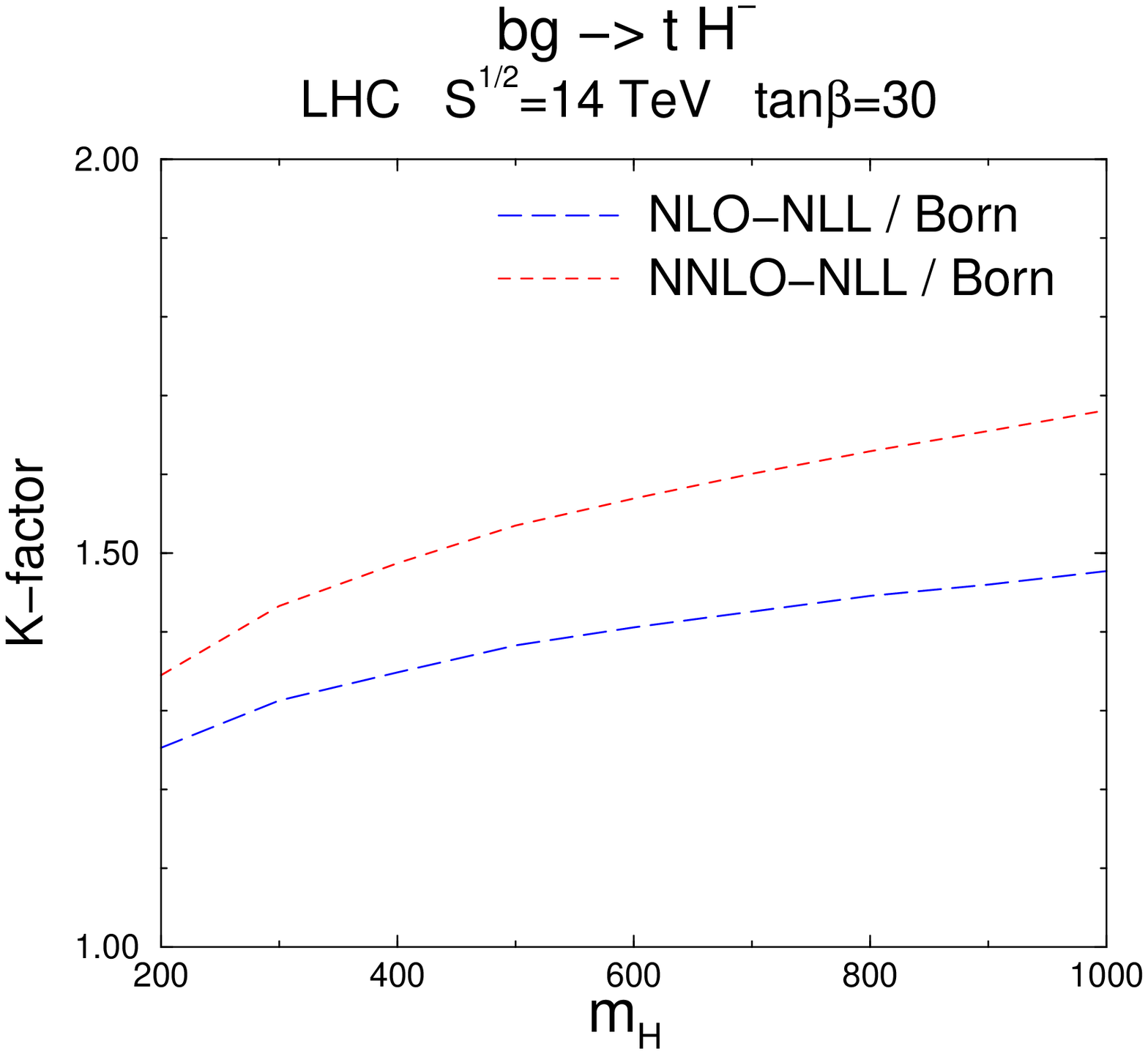}
\caption[]{Charged Higgs production at the LHC.
Left: The total cross section. Right: The $K$-factors.}
\label{chiggs}
\end{figure}

In Fig. \ref{chiggs} we show results for charged Higgs production
in $pp$ collisions at the LHC.
We set $\mu_F=\mu_R=m_H$ and use the MRST2002 NNLO ~\cite{MRST2002} 
parton densities. We keep $m_b$ non-zero only in the coupling.
In the left frame of Fig. \ref{chiggs} we plot the cross section for charged 
Higgs production with $\sqrt{S}=14$ TeV versus 
the charged Higgs mass. We use $m_t=175$ GeV, $m_b=4.5$ GeV, 
and $\tan \beta=30$, where $\tan \beta=v_2/v_1$ is the ratio
of the vacuum expectation values of the two Higgs doublets 
in the MSSM.
The Born, NLO-NLL, and NNLO-NNLL results are shown.
Both the NLO and the NNLO threshold corrections are important
as can be more clearly seen in the right frame of Fig. 1 where
we plot the $K$-factors, i.e. the ratios of the NLO-NLL over Born
and the NNLO-NLL over Born cross sections. As expected, the corrections
increase for larger charged Higgs mass at fixed $\sqrt{S}$, since then 
we get closer to the threshold region. Finally, we note that the cross 
section for ${\bar b} g \longrightarrow {\bar t} H^+$ is the same 
as for $bg \longrightarrow t H^-$.

\mysection{Conclusions}

I have reviewed recent calculations of soft-gluon corrections
for processes in hadron-hadron and lepton-hadron collisions.
A unified approach and master formulas that allow the
explicit calculation of next-to-next-to-leading order soft
and virtual corrections for arbitrary processes
have been presented. 
I have also presented phenomenological applications of the 
unified formalism to top quark pair production, 
FCNC top production via anomalous 
$tqV$ couplings, $W$-boson hadroproduction
at large transverse momentum, and charged Higgs production.
In all cases the soft corrections are significant and important,
and they greatly decrease the dependence of the cross sections on 
factorization and renormalization scales.

\mysection*{Acknowledgments}

I would like to thank Sasha Belyaev, Jeff Owens, Agustin Sabio Vera, 
and Ramona Vogt for fruitful collaborations.
The author's research has been supported by a Marie Curie Fellowship of 
the European Community programme ``Improving Human Research Potential'' 
under contract number HPMF-CT-2001-01221.

\end{document}